# Identification of bifurcations from observations of noisy biological oscillators

Joshua D. Salvi, Dáibhid Ó Maoiléidigh, and A. J. Hudspeth[†]

Howard Hughes Medical Institute and Laboratory of Sensory Neuroscience, The Rockefeller University, 1230 York Avenue, New York, New York, 10065, USA

Running Title: Identification of noisy bifurcations

[†] Correspondence: hudspaj@rockefeller.edu

# Abstract

Hair bundles are biological oscillators that actively transduce mechanical stimuli into electrical signals in the auditory, vestibular, and lateral-line systems of vertebrates. A bundle's function can be explained in part by its operation near a particular type of bifurcation, a qualitative change in behavior. By operating near different varieties of bifurcation, the bundle responds best to disparate classes of stimuli. We show how to determine the identity of and proximity to distinct bifurcations despite the presence of substantial environmental noise. Using an improved mechanical-load clamp to coerce a hair bundle to traverse different bifurcations, we find that a bundle operates within at least two functional regimes. When coupled to a high-stiffness load, a bundle functions near a supercritical Hopf bifurcation, in which case it responds best to sinusoidal stimuli such as those detected by an auditory organ. When the load stiffness is low, a bundle instead resides close to a subcritical Hopf bifurcation and achieves a graded frequency response—a continuous change in the rate but not the amplitude of spiking in response to changes in the offset force—a behavior useful in a vestibular organ. The mechanical load *in vivo* might therefore control a hair bundle's responsiveness for effective operation in a particular receptor organ. Our results provide direct experimental evidence for the existence of distinct bifurcations associated with a noisy biological oscillator and demonstrate a general strategy for bifurcation analysis based on observations of any noisy system.

# Introduction

A bifurcation occurs when a quantitative change in the value of some parameter—a control parameter—induces a qualitative change in the behavior of a system. Bifurcations are often encountered in theoretical and experimental systems in physics, chemistry, biology, medicine, economics, and climatology (1-4). All systems operating near a given type of bifurcation exhibit similar dynamics (5), and systems operating near different bifurcations can exhibit distinct behaviors. Identifying a bifurcation therefore reveals generic properties of a complex system and permits prediction of that system's function when it operates near the bifurcation.

The activity of spiking neurons, for example, can be segregated into at least two classes of excitability with distinct patterns in the frequency of spiking that are attributed to operation near different types of bifurcation (6-8). Class 1 neurons display a continuous change in spike frequency in response to changes in stimulus current, a behavior commonly ascribed to their operation near a saddle-node on invariant cycle (SNIC) bifurcation (9). Class 2 neurons, by contrast, exhibit a discontinuous jump in frequency as a function of the applied current, a behavior consistent with their crossing a Hopf bifurcation (10). By identifying the type of bifurcation, one can assess generic features of neuronal excitability in different neuronal populations.

Bifurcation analysis is usually conducted on purely mathematical stereotypes or on simplified representations of real-world systems for which the effects of noise are small. In certain experimentally accessible systems, however, noise plays a significant and irreducible role in shaping the dynamics. Among these systems is the hair cell, the sensory receptor of the auditory, vestibular, and lateral-line sensory systems of vertebrate animals (11). A hair cell detects mechanical signals derived from sounds, accelerations, and water movements, transducing them into electrical signals that are transmitted to the brain. This detection is achieved through the motion of a mechanosensitive organelle—the hair bundle—that projects from the hair cell's apical surface and transduces mechanical input into electrical output in the cell body. For small-magnitude stimuli the hair bundle operates in an environment dominated by noise. In fact, the sensitivity of our hearing is limited by the clattering of air molecules against the eardrum and the rattling of water molecules within the cochlea (12, 13). Thermal noise also plays an important role in setting the sensitivity of vestibular organs (14).

Auditory organs deal with noise by employing an active process, a metabolically powered mechanism that enhances their sensitivity (15). Vestibular systems might also employ the active process to improve their effectiveness. The effects of the active process depend on the values of a sensory organ's parameters (16). For example, it has been hypothesized that auditory hair bundles achieve enhanced frequency selectivity and a broadened dynamic range in response to periodic stimuli when they are poised near the onset of spontaneous oscillation, that is, when they operate close to a supercritical Hopf bifurcation (17-19). Observations of individual bundles support this hypothesis (20). Hair bundles have also been observed to attain a graded frequency response with changes in static offsets, which could occur if hair bundles operate near a SNIC bifurcation or—as we argue below—a subcritical Hopf bifurcation (20-22). We propose that this graded response allows hair bundles to function as static force detectors, which would be of use in a vestibular system. By operating near particular bifurcations, a mechanically active hair bundle can therefore be useful in different mechanosensory systems.

A bundle's state diagram depicts its behavior for different operating points defined by the values of the system's control parameters (16, 20). These control parameters include the mechanical loads imposed on individual hair bundles by accessory structures *in vivo*, such as the tectorial membrane coupled to the bundles of outer hair cells in the mammalian cochlea or the otolithic membrane coupled to hair cells of the utriculus and sacculus. Each structure loads the bundle with a mass, a drag, a stiffness, and an offset force; the combination allows a bundle to operate near the type of bifurcation that is suitable for its sensory role.

Although biophysical experiments have identified several of the mechanisms that underlie the active process (15), the noisy environment of a hair bundle complicates the identification of the relevant bifurcations. In particular, noise blurs the boundaries between the dynamical regimes in the bundle's state diagram and conceals the characteristics of various bifurcations. There are several types of bifurcation in which a quiescent system becomes self-oscillatory. In the absence of noise, a system crossing either a SNIC or a Hopf bifurcation exhibits respectively a continuous rise or a discontinuous jump in the frequency of spontaneous oscillation (5-7). Changes in the amplitude of oscillation with the adjustment of a parameter further distinguish two types of Hopf bifurcation: a gradual rise in amplitude corresponds to a supercritical Hopf bifurcation and a discontinuous jump signals a subcritical Hopf bifurcation (7). In the presence of noise, however, sharp transitions in the amplitude and frequency of spontaneous oscillation are blunted and become difficult to differentiate from their gradual counterparts, complicating both the localization and the identification of bifurcations. We have overcome this challenge by employing a battery of statistical tests to locate and identify bifurcations from experimental observations of a noisy system.

Deterministic bifurcations, those defined in the absence of noise, occur at well-specified parameter values. For example, a deterministic system possessing a supercritical Hopf bifurcation is quiescent on one side of the bifurcation and oscillatory on the other. Although the location of this bifurcation can only be estimated from observations, we can nonetheless identify its quiescent and oscillatory sides to define the bifurcation empirically. By analogy with the deterministic case, we have defined the location of an empirical bifurcation for a noisy system as the parameter value at which spontaneous oscillations can be reliably detected from observations. The quiescent side of this empirical bifurcation covers the range of parameter values for which oscillations cannot be detected with sufficient statistical certainty. In this manner we have delineated the oscillatory and quiescent regions near supercritical Hopf, subcritical Hopf, and SNIC bifurcations from noisy data.

We have explored various metrics that capture the distinguishing features of systems crossing different bifurcations. To do so, we have compared simulations of noisy systems near various bifurcations with experimental observations of noisy hair-bundle motion. Based on these comparisons, we have ascertained which metrics best characterize the identities and locations of bifurcations near which hair bundles operate. By employing these metrics to assess the behavior of individual hair bundles, we have established that a hair bundle can operate near at least two types of bifurcation. We have then associated the bundle's operation near these bifurcations with the functions of auditory and vestibular receptor organs. Our methodology not only furthers an understanding of hair-bundle dynamics but also provides a general strategy for analyzing empirical observations of noisy dynamical systems.

## Materials and Methods

All experiments were performed on spontaneously active hair bundles from the saccular maculae of adult bullfrogs, *Rana catesbeiana* (20).

**Supporting Material.** We have included detailed Supporting Material to accompany this paper. Although the entirety of the document is not required for a general understanding of our findings, we prepared the Supporting Material so that it can serve as a self-contained guide for those who wish to apply our approach to other systems.

**Mechanical-load clamp.** We employed a feedback system with a real-time interface to control the load stiffness and constant force applied to individual hair bundles (23). Using a glass fiber as a model hair bundle, we confirmed that the clamp achieves this performance with high precision and accuracy (Fig. S1). For a complete description of the mechanical-load clamp, see Supporting Material Section A.

**State-diagram mapping.** To calculate the amplitude and frequency of hair-bundle motion in response to various combinations of load stiffness and constant force, we employed both described methods (20) and a peak-detection algorithm (24). Operating points for which the bundle's position histogram displayed at least two clear maxima, as signaled by a Hartigans' dip statistic that exceeded 0.01 with $p < 10^{-3}$, were classified as oscillatory (25). All other operating points were deemed quiescent. The Hartigans' dip statistic is described in more detail below, and an expanded explanation of state-diagram calculations can be found in Supporting Material Section B.

**Time-series analysis.** We employed a battery of statistical tests to assess the behavior of experimentally observed hair bundles and of simulated systems. We included simple tests that limited the number of manually selected parameters. Our approach requires only noisy time-series data and does not necessitate stimulation. These features enhance the versatility of the method, permitting its use in systems for which stimulation is difficult or impossible. For a complete description of these metrics, see Supporting Material Section C.

A bifurcation occurs when a system undergoes a qualitative change in behavior. For example, the system may transition from a domain of quiescence to one of spontaneous oscillation. We first sought to estimate the location of a bifurcation by detecting the onset of spontaneous oscillations across a range of control-parameter values. A system that oscillates spontaneously displays a position distribution with more than one peak (20, 26, 27). Although previous studies measured the distance between peaks in the distribution to describe this phenomenon (27, 28), we instead employed Hartigans' dip statistic to measure the modality of a distribution (25). Not only is the dip statistic an inferential measure of modality with an associated *p*-value, it is also less biased by skew and sample size relative to other metrics (29). Whereas a small value of the dip statistic corresponds to a position distribution with one peak, a large value reflects a distribution that has more than one peak. We therefore defined a position distribution possessing a large value of the dip statistic with a correspondingly small *p*-value as an indicator of spontaneous oscillation, and a small value of the dip statistic as an indicator of quiescence.

The behavior of a system is typically recorded using only a single observable, such as a bundle's position in time. However, in many instances a system is described by more than one variable and the system's dynamics occupies a space of many variables called a phase space. A dynamical trajectory in phase space possesses information that is not apparent from a single-variable time series. We therefore wished to reconstruct a two-dimensional phase-space representation of hair-bundle dynamics based only on observations of its displacement. Because many methods of phase-space reconstruction require the manual selection or empirical estimation of several parameter values (30-33), we focused instead on a simpler alternative. The Hilbert transform of a signal is the imaginary part of its analytical representation (34). The joint probability distribution of a bundle's real-valued position and the Hilbert transform of that position—an analytic distribution—is a distribution over a system's phase space that can be calculated without the need for complicated analysis and challenging parameter selection. The analytic distribution appears ring-shaped when a system displays limit-cycle oscillations, which correspond to stable, closed-loop trajectories in phase space. Near quiescent fixed points the distribution instead exhibits an enhanced, disk-shaped density. This representation is similar to that used to analyze spontaneous otoacoustic emissions (35).

We next estimated the frequency, amplitude, and regularity of spontaneous oscillations with a peak-detection algorithm that allows the detection in a time series of all peaks and troughs that cross defined thresholds (24). The method permits calculation of a system's amplitude and frequency of oscillation for time series with substantial noise, short durations, and non-sinusoidal waveforms. For each peak-detection threshold, the frequency is defined as the inverse of the mean time interval between peaks, and the amplitude is calculated as half of the mean difference in position between each peak and trough.

We did not employ spectral methods to determine the location or identity of a bifurcation. In the presence of noise, both a quiescent resonant system and a limit-cycle system exhibit peaks in their Fourier spectra (36). A qualitative change in spectra as this noisy system crosses a bifurcation does not exist. In contrast, a system's position distribution exhibits a qualitative change when the system crosses a bifurcation. Moreover, because many of our signals were subjected to substantial noise and displayed non-sinusoidal waveforms, the Fourier frequency and amplitude of oscillation often fluctuated considerably as the value of a control parameter was changed, and consequently this approach did not yield a clear estimate of a bifurcation's location that was consistent across the systems we studied (Figs. S6-S8). The peak-detection algorithm also performed more reliably than spectral analysis in estimating the amplitude and frequency of noise-induced spikes (Fig. S9).

As a system transitions from a regime dominated by limit-cycle oscillations to one dominated by noise, that system's spontaneous motion becomes more irregular. The distribution of interpeak time intervals, the times between neighboring peaks, correspondingly broadens. We therefore assessed the regularity of a system's oscillations by calculating the coefficient of variation, defined as the ratio of this distribution's standard deviation to its mean. As a system's oscillations become increasingly irregular, the metric grows and thus traces the transition from motion governed by limit-cycle oscillations to that dominated by noise. The regularity of oscillations determined using the coefficient of variation implies that multiple peaks in the position histogram stem from limit-cycle oscillations rather than from noise-induced switching between stable states. For example, a system crossing a saddle-node, or fold, bifurcation may possess a stable state on one side of the bifurcation and two stable states on the other. This system can stochastically switch between two states in the bistable regime. We found that the coefficient of variation for a noisy system transitioning from monostability to bistability is large for all control-parameter values, in contrast with the smaller coefficients of variation displayed by systems displaying limit-cycle oscillations (Fig. S12).

Finally, we estimated the location of a bifurcation using an information-theoretical metric. Upon crossing certain bifurcations, a system exhibits limit-cycle oscillations that appear ring-shaped in that system's analytic distribution. To quantify this phenomenon, we estimated the mutual information between the real and imaginary parts of the system's analytic signal, which we entitle the analytic information. The analytic information resembles the time-delayed mutual information, a nonlinear measure of temporal correlation in a signal (37). As opposed to a shift of the signal in *time*, however, the analytic information is calculated from a real-valued signal and that signal shifted in *phase*. In the presence of a limit cycle the analytic information is large; in the absence of a cycle its value is small. For example, the analytic information approaches zero for a narrow-band Gaussian-distributed noise sequence as its length increases (Fig. S10). The information increases as a system crosses a bifurcation and begins to oscillate spontaneously. Together with the dip statistic, the analytic information therefore serves as an indicator of the onset of limit-cycle oscillations.

**Numerical simulations.** A system operating near a bifurcation can be described by the normal form corresponding to the type of bifurcation. The normal form is a standard mathematical expression that captures the generic features of any system operating near that kind of bifurcation. We performed simulations of the normal forms for different bifurcations, in which we included various levels of additive noise for each variable. We simulated the normal forms of the supercritical Hopf, the subcritical Hopf, and the SNIC bifurcations. We also simulated a model of hair-bundle dynamics with additive noise (16). We refer to all these simulations as stochastic simulations throughout the paper. All simulations were conducted in MATLAB (R2014a, 8.3.0.532) with the Euler-Murayama method.

Because the simulations of the normal forms and the hair-bundle model have no meaningful dimensions in position and time, the results of the simulations are reported without units. In keeping with the convention in the dynamical-systems literature, we orient the abscissa of every plot such that the quiescent regime occurs to the left and the oscillatory regime to the right of each bifurcation. For a complete description of these simulations, see Supporting Material Sections D and E.

## Results

The hair bundles of various receptor organs confront a variety of physical loads (15). Both modeling and experiments show that adjusting the mechanical load qualitatively changes a hair bundle's behavior (16, 20). Our previous work was limited, however, by the stability and precision of the clamp system used to apply the load on a hair bundle. These limitations impeded determination of the location and identity of bifurcations displayed by a bundle. After improving the mechanical-load clamp to overcome these restrictions (Figs. S1-S3), we have probed the nature of bifurcations more systematically than heretofore.

### The hair bundle's state diagram

Changing the load stiffness and constant force applied to an individual hair bundle yields a map of the bundle's behavior—a state diagram—for combinations of these two control parameters (Fig. 1A). We predicted earlier that a hair bundle oscillates spontaneously within a region bounded by subcritical and supercritical Hopf bifurcations and exhibits bistability in a regime bounded by lines of fold bifurcations (16). Within the oscillatory regime, we also expect the bundle's spontaneous motion to fall in amplitude and rise in frequency with an increase in load stiffness. In the present study, we have explored a bundle's behavior within different locales of its state diagram and identified the types of bifurcation near which the bundle can operate.

Using the improved load clamp, we determined the behavior of a bundle as a function of its mechanical load. In agreement with the theoretical expectation, the load stiffness controlled the character of the bundle's spontaneous oscillations (Fig. 1B). When the stiffness was small, the bundle oscillated at low frequency and with a high amplitude and in some cases exhibited mixed-mode oscillations. Increasing the load stiffness caused the bundle's oscillations to increase in frequency and decrease in amplitude until they vanished altogether.

A spontaneously oscillating hair bundle displays a distribution of positions with more than one peak, whereas the position distribution of a quiescent bundle has only a single peak (20, 26, 27). We therefore classified the bundle's behavior as either oscillatory or quiescent based on its position histogram. To do so, we employed Hartigans' dip statistic, for which a large value corresponds to a multimodal position distribution generated by a spontaneously oscillating bundle and a small value arises from a unimodal position distribution produced by a quiescent bundle (25). This metric revealed an ovoid oscillatory regime surrounded by a domain of quiescence (Figs. 1C-D and S4). The boundary between these two regimes is expected to correspond to lines of subcritical and supercritical Hopf bifurcations (16, 20). Improvements in data acquisition and analysis allowed us to construct the experimental state diagram with greater stability and a larger signal-to-noise ratio over this range of control parameters than heretofore, eliminating the need for additional statistics to delineate the locus of oscillation (20). The frequency of spontaneous oscillation rose and the amplitude fell with an increase in load stiffness. With an increase in the constant force, however, both the amplitude and the frequency declined (Fig. 1E). These correlations accord with theoretical predictions and prior experimental investigation of the effects of load stiffness on a bundle's behavior (16, 20).

Although the experimental state diagram depicted the boundary of and patterns within a regime of spontaneous oscillation, the identity of the bifurcations that define the boundary has not been determined definitively. Because a bundle's operation near a particular bifurcation can in part dictate its mechanosensory function, we developed a strategy for the identification of bifurcations in experimental systems with substantial noise.

**A supercritical Hopf bifurcation for high stiffness loads**

We first assessed whether a hair bundle in the high-stiffness regime can operate near a supercritical Hopf bifurcation. To do so, we compared simulations of a noisy system crossing a supercritical Hopf bifurcation, described by the bifurcation's normal-form equation, with experimental observations of a hair bundle subjected to a high load stiffness and increasing values of constant force (Fig. 2). Both the model system and the experimentally observed hair bundle oscillated spontaneously with amplitudes that grew with respectively an increase in the control parameter or a decrease in the bundle's constant force (Figs. 2A, H and S5A). When the amplitude of oscillation exceeded the noise, the position histogram displayed two distinct peaks. The analytic distribution, a two-dimensional representation of the bundle's motion, formed a loop corresponding to a limit cycle. An increase in the control parameter or a decrease in the bundle's constant force caused the diameter of the loop to grow in concert with a rise in the amplitude of oscillation.

We determined the location of an empirical bifurcation as the point at which the dip statistic assumed a statistically significant value. Our estimate of the noisy bifurcation's location resided on the oscillatory side of a deterministic supercritical Hopf bifurcation (Fig. 2B). For an active hair bundle, we found a sharp transition between small and large values of the dip statistic, corresponding to a bifurcation defining the boundary between a quiescent and an oscillatory regime (Fig. 2I).

Although a system's amplitude and frequency of motion are typically calculated from its Fourier transform, the Fourier amplitude and frequency can be uninformative for determining a bifurcation's location when noise is substantial, when the signals are brief, or when oscillations deviate substantially from pure sinusoids. We illustrate this problem with the Fourier transform both for noisy simulations and for experimental observations of hair-bundle motion (Supporting Material Section C.7). For an oscillator operating far from a bifurcation, spectral analysis performs well. Near a bifurcation and in the presence of noise, however, detection of the bifurcation using the Fourier amplitude and frequency becomes difficult. We show that the amplitude and frequency calculated with the Fourier transform fluctuate considerably as a system's control parameter is changed (Fig. S8). Furthermore, spectral methods do not reliably capture the amplitude and frequency for a system that exhibits noise-induced spikes (Fig. S9). Consequently, this method can fail to detect bifurcations in systems dominated by noise.

We therefore characterized the amplitude of spontaneous oscillation using two other metrics. First, we calculated the root-mean-square (RMS) magnitude. For both the simulation and the experimentally observed hair bundle, the RMS magnitude rose gradually as the operating point was moved toward the region of spontaneous oscillation (Fig. 2C, J). However, both the amplitude and the frequency of oscillation can affect the RMS magnitude. For example, constant-amplitude spikes that become less frequent correspond to a declining RMS magnitude even though their amplitude does not change. To circumvent this issue, we additionally determined the amplitude of oscillation from a peak-detection algorithm that found the local maxima (peaks) and minima (troughs) separated by a threshold distance. Our peak-detection algorithm accurately estimated the amplitude and frequency of oscillation (Supplementary Material Section C.7). We defined the amplitude as half of the average distance between the position of a peak and its neighboring trough. As the control parameter grew or the bundle's constant force shrank, the amplitude rose sharply and then more gradually regardless of the peak-detection threshold (Fig. 2D, K). The amplitudes calculated from the peak-detection algorithm were sensitive to the selected peak-detection threshold. For both the simulated and the experimentally observed time series, the amplitude curve shifted toward the oscillatory region as the threshold rose.

We then used the same peak-detection algorithm to calculate the frequency of oscillation, which we defined as the inverse of the mean interval between successive peaks. As noted for the amplitude relations, the frequencies calculated with this method were sensitive to changes in the peak-detection threshold. Increasing the value of the peak-detection threshold shifted the frequency curve farther into the oscillatory side of the bifurcation for both the noisy simulations and the experimental observations (Fig. 2E, L).

As a system transitions from a regime dominated by large-amplitude oscillations to one dominated by noise, its spontaneous motion becomes increasingly irregular. We therefore quantified the variability of a system's spontaneous motion by the coefficient of variation for the system's interpeak time intervals: large and small coefficients correspond respectively to irregular and regular oscillations. For all peak-detection thresholds, this metric fell as the systems' operating points moved farther into the oscillatory regime, indicating less variation in the interpeak intervals as the system's motion became dominated by limit-cycle oscillations (Fig. 2F, M). Increasing the value of the peak-detection threshold shifted the location at which the coefficient of variation crossed a threshold of 0.5 farther into the oscillatory regime for both a noisy system crossing a supercritical Hopf bifurcation and for an experimentally observed hair bundle. As with the amplitude and frequency of motion, the coefficient of variation depended on the peak-detection threshold. The point at which the coefficient crossed 0.5 moved toward the oscillatory regime with an increase in the value of the peak-detection threshold.

A noisy bistable or multistable system exhibits a position histogram with more than one peak and consequently a large value for the dip statistic. The dip statistic alone therefore cannot distinguish limit-cycle oscillations from noise-induced switching between stable states. However, position fluctuations are much less coherent for bistable and multistable systems than for a limit-cycle oscillator. We can therefore use the coefficient of variation to determine whether a system exhibits limit-cycle behavior or whether it displays noised-induced switching between stable states. The coefficient of variation never falls below 0.5 for a bistable or multistable system, but it can approach zero for a limit-cycle oscillator (Supporting Material Section D.6).

Finally, we sought to pinpoint the location of a bifurcation and to determine its identity with another metric. The analytic information, defined as the mutual information between the real and imaginary parts of the system's analytic signal, approaches zero for normally distributed noise and grows as limit-cycle oscillations emerge. The analytic information rose gradually with an increase in the control parameter in the simulation and with a decrease in the bundle's constant force in the experiment (Fig. 2G, N). Although the gradual rise in the analytic information failed to identify the exact location of a bifurcation, the trends for both the model and experiment displayed strong similarity. Taken together, the striking agreement between simulations and experimental observations implies that a hair bundle subjected to a large load stiffness can traverse a supercritical Hopf bifurcation as the constant force is changed.

To confirm that the high-stiffness boundary between the oscillatory and quiescent regions constitutes a line of supercritical Hopf bifurcations, we subjected both a model hair bundle and an experimentally observed one to a constant force of zero and decreasing values of the load stiffness, a regime for which the model bundle is known to cross a supercritical Hopf bifurcation. We then employed the same battery of tests as before to assess the similarities between the time series of the simulated and experimentally observed hair bundles (Fig. 3). All panels in Figure 3 display the same metrics as the corresponding panels in Figure 2 but for different time series.

In agreement with the generic features of a noisy system crossing a supercritical Hopf bifurcation (Fig. 2A), both the simulated and the experimentally observed hair bundle displayed spontaneous oscillations whose amplitude grew with a decrease in stiffness (Figs. 3A, H and S5B). Their analytic distributions also displayed limit cycles that correspondingly increased in diameter. Consistent with that of a noisy system near a supercritical Hopf bifurcation (Fig. 2B), the simulated bundle's position histogram became bimodal on only the oscillatory side of the deterministic bifurcation (Fig. 3B). We observed a clear transition at which simulated and experimentally observed hair bundles became oscillatory (Fig. 3B, I).

Both the RMS magnitude (Fig. 3C, J) and the amplitude (Fig. 3D, K) of spontaneous oscillation rose gradually with a decrease in stiffness; this pattern accorded with that of a noisy system crossing a supercritical Hopf bifurcation (Fig. 2C-D). The frequency of oscillation, however, followed a trend that differed from our simulations of generic dynamics near that bifurcation. As the stiffness decreased from its greatest value, the frequency of oscillation first rose for both the simulated and the experimentally observed hair bundle in agreement with a quiescent system's approach to a supercritical Hopf bifurcation (Fig. 3E, L). Further decreases in stiffness, however, caused the frequency to achieve a maximal value and subsequently to decline as the operating point moved beyond the range of influence of the bifurcation. Although this pattern in frequency differs from the generic behavior of a system operating near a supercritical Hopf bifurcation, it accords with the specific behavior predicted for hair-bundle dynamics (Fig. 1A). The coefficient of variation was once again sensitive to changes in the peak-detection threshold, consistent with the generic behavior of a system operating near a supercritical Hopf bifurcation (Fig. 2E-F). For both the simulated and the experimentally observed bundles, the coefficient of variation crossed 0.5 at smaller stiffness values when the value of the peak-detection threshold was larger (Fig. 3E-F, L-M). Finally, the analytic information rose gradually as the systems' operating points advanced into the oscillatory regime (Fig. 3G, N), in agreement with that of a system crossing a supercritical Hopf bifurcation (Fig. 2G, N).

All the observations of hair-bundle behavior agreed with stochastic simulations of the normal form of a supercritical Hopf bifurcation and of a model of hair-bundle dynamics. Taken together, these data strongly support the prediction that for an active hair bundle the boundary of oscillation for large stiffnesses is a line of supercritical Hopf bifurcations (16, 20).

**A subcritical Hopf bifurcation for low stiffness loads**

We predicted that a hair bundle subjected to a small load stiffness operates near a subcritical Hopf bifurcation (Fig. 1A), whereas an alternate analysis suggested that the bundle approaches a SNIC bifurcation (22, 38). To evaluate these alternatives, we used the methodology discussed above to compare experimental observations of a hair bundle subjected to a small load stiffness to those of noisy systems operating near either a subcritical Hopf or a SNIC bifurcation. All panels in Figures 4 and 5 correspond to the metrics displayed in Figures 2 and 3 but for different sets of simulations and experimental observations.

When operating on the oscillatory side of either a subcritical Hopf or a SNIC bifurcation, the noisy systems exhibited large-amplitude oscillations (Fig. 4A, H) that accorded with those of an experimentally observed hair bundle subjected to a small stiffness (Fig. 4O). On the other side of a subcritical Hopf bifurcation, however, this system displayed noise-induced bursts of high-amplitude oscillations (Fig. 4A). A system near a SNIC bifurcation instead exhibited noise-induced spikes with a frequency that fell with a decrease in the control parameter (Fig. 4H). The bursting behavior near a subcritical Hopf bifurcation occurred when a limit cycle coexisted with a stable point at its center, as evidenced by an analytic distribution possessing an enhanced density within a loop (Figs. 4A and S5C). Fluctuations induced transitions between the stable point and the limit cycle, resulting in bursts of noisy oscillations. Bursting occurred only for parameter values within the region of coexistence that fell between the subcritical Hopf bifurcation and a saddle-node of limit cycles bifurcation. The latter bifurcation, at which a stable and an unstable limit cycle collide and annihilate one another, is a consequence of operation near a subcritical Hopf bifurcation. The spiking behavior exhibited by a system near a SNIC bifurcation, by contrast, engendered a locus of high probability along one part of a cycle (Figs. 4H and S5C). The bundle's spontaneous motion most closely resembled the spiking behavior and analytic distributions of a noisy system proximal to a SNIC bifurcation (Figs. 4O and S5C).

Whereas the dip statistic indicated noisy oscillations solely on the oscillatory side of a deterministic supercritical Hopf bifurcation (Fig. 2B), it instead evidenced noisy oscillations not only on the deterministically oscillatory side but also within the coexistence region of a deterministic subcritical Hopf bifurcation (Fig. 4B). The dip statistic also detected oscillations on the oscillatory side of a deterministic SNIC bifurcation (Fig. 4I). The dip statistic for an experimentally observed hair bundle clearly delineated the boundary between a quiescent regime and an oscillatory one, again demonstrating the utility of this metric in pinpointing a bifurcation's location (Fig. 4P).

All the other metrics were qualitatively indistinguishable between noisy systems operating near a subcritical Hopf or a SNIC bifurcation, with behaviors from both systems resembling those of an experimentally observed hair bundle subjected to a low stiffness. However, these behaviors were distinct from those of a system close to a supercritical Hopf bifurcation (Figs. 2 and 3). In all cases, the RMS magnitude increased abruptly near a bifurcation (Fig. 4C, J, Q). The amplitude rose suddenly before a bifurcation and then grew gradually or remained constant (Fig. 4D, K, R). Unlike the pattern of a system operating near a supercritical Hopf bifurcation, the rise in the frequency of spontaneous oscillation and the fall in the coefficient of variation with an increase in control parameter were insensitive to the value of the peak-detection threshold (Fig. 4E-F, L-M, S-T). Finally, the analytic information rose abruptly near the bifurcation (Fig. 4G, N, U), in contrast with the gradual rise displayed by a system near a supercritical Hopf bifurcation (Fig. 2G). These data together demonstrate that, for a low stiffness, a hair bundle can cross either a subcritical Hopf or a SNIC bifurcation but not a supercritical Hopf bifurcation.

We next assessed whether the bursting behavior of a system near a subcritical Hopf bifurcation disqualifies this bifurcation as that exhibited by observed hair bundles at low stiffnesses. Using the same battery of tests, we compared experimental observations of another hair bundle with stochastic simulations of a model bundle that is known to operate near a subcritical Hopf bifurcation (Figs. 1A, 5). In the presence of noise, the simulated bundle exhibited downward excursions resembling the spikes displayed by both of the experimentally observed bundles at low stiffnesses (Figs. 4O, 5A, H, and S5D). Unlike a system displaying bursting near a subcritical Hopf bifurcation (Fig. 4B), the simulated bundle showed noise-induced spikes well beyond the saddle-node of limit cycles bifurcation (Fig. 5B). The bundle displayed a graded frequency response: the spiking frequency fell with an increase in the bundle's constant force. This change was accompanied by a locus of increasing probability in the analytic distribution that rested upon one part of a large-amplitude limit cycle. Spikes may therefore result either from the asymmetric dynamics associated with a SNIC bifurcation or from the specific asymmetry captured by the hair-bundle model near a subcritical Hopf bifurcation, but lacking in the normal form of a subcritical Hopf bifurcation. All metrics showed strong qualitative agreement between the model bundle operating near a subcritical Hopf bifurcation, a SNIC bifurcation, and both experimentally observed hair bundles (Figs. 4 and 5). The changes in the metrics with control parameter were once again distinct from those of a noisy system traversing a supercritical Hopf bifurcation.

## Discussion

A hair bundle's function can be dictated in part by its operation near a particular bifurcation. We have identified the types and locations of bifurcations from experimental observations of noisy hair bundles. By employing several metrics to compare models with experimental observations, we have analyzed the bifurcation structure of experimentally observed hair bundles operating in a noisy environment and confirmed the predictions of a qualitative model of hair-bundle dynamics.

Our model makes no assumptions about the temporal or spatial scales of a bundle's dynamics and requires only two properties of hair bundles: their nonlinear stiffness and adaptation to stimulation (15, 39-42). Any actual bundle or hair-bundle model possessing these features exhibits the same state diagram as our model and thus qualitatively similar responses to stimulation (16). Moreover, we previously used a quantitative model of hair-bundle mechanics with physical parameters plausible for the mammalian cochlea to demonstrate that hair-bundle activity is likely essential for the mammalian auditory system to achieve great sensitivity and sharp frequency selectivity in response to high-frequency periodic stimuli (43).

Guided by the qualitative model, we found that a single hair bundle can operate near more than one type of bifurcation, depending on its mechanical load. Although it can be argued that noise introduces new bifurcations and changes the character of existing ones (44), our observations accord well with simulations of systems crossing bifurcations in the presence of noise.

**Proximity to a bifurcation**

To understand certain dynamical systems one must determine whether they can operate near bifurcations. We employed a battery of quantitative metrics to isolate the location of such bifurcations as a function of a control parameter. Of these techniques, the Hartigans' dip statistic offered multiple advantages when used to determine the modality of a bundle's position distribution. First, the dip statistic is an inferential metric, providing an associated *p*-value that allowed us to estimate the bifurcation's position consistently across all the data sets. Second, the dip statistic is less sensitive to sample size and skew than other measures of a distribution's modality (29).

A phenomenological bifurcation occurs when the probability distribution of a system's state, including but not limited to its position distribution, exhibits a qualitative change; for example, the distribution's modality may change (44). A third benefit of using the dip statistic is that it identified phenomenological bifurcations associated with changes in the position distribution. Although this approach might have missed modality changes in the bundle's state distribution associated with unobserved variables, such as the hair cell's transduction current, it clearly defined a boundary between an oscillatory and a quiescent regime based only on experimental observations of the bundle's motion. Although we found that noise introduced an unavoidable bias in the estimation of a deterministic bifurcation's position, we note that phenomenological bifurcations can occur at values of a control parameter that differ from those in the deterministic cases.

The coefficient of variation for the values of a system's state variables in a time series was previously employed to experimentally estimate the locations of Hopf bifurcations in a predator-prey system, in which the bifurcation's location was defined by the transition from a small to large value of the coefficient corresponding to the onset of spontaneous oscillations (45). By using the magnitude of the state variables rather than the time intervals between events, this coefficient of variation captures a different aspect of a dynamical system's behavior than the coefficient of variation we employ here. Because this coefficient of variation does not describe the regularity of a system's oscillations or spiking, it cannot distinguish between noisy switching in a bistable system and limit-cycle oscillations. As the mean values of the variables we study is often zero, as is the case for the Hopf normal form, it is not possible to calculate a coefficient of variation. Moreover, a change in this metric with control parameter could arise from a change in a variable's mean value rather than from a system crossing a Hopf bifurcation. Nonetheless, this metric is likely useful for bifurcation analysis of some systems especially if it were to be combined with some of the measures we utilize here.

**Identity of a bifurcation**

We developed a protocol that permits the identification of bifurcations solely on the basis of noisy time series. This diagnostic method has several appealing features. Although the approach requires data from a range of operating points near a bifurcation, it does not require external stimulation. In the study of climate change, finance, and geophysics (3, 4), among other disciplines, stimulation may be difficult or even impossible. Moreover, the method performs well at high noise levels and relies on few if any choices by the experimenter.

Although the dip statistic can be employed to locate a bifurcation, it cannot be used to distinguish between types of bifurcation. To identify the bifurcation near which a system operates, we applied five additional metrics, each of which captures a different feature of the system's behavior (Fig. 6). These metrics and the analytical distributions allow us to distinguish supercritical Hopf bifurcations from subcritical Hopf and SNIC bifurcations. Near a supercritical Hopf bifurcation, the RMS magnitude grows more slowly with the control parameter and the oscillation frequency and coefficient of variation are more dependent on the peak-detection threshold than for the subcritical Hopf and SNIC bifurcations. In addition, the analytical distribution often evidences a fixed point surrounded by a limit cycle when a bundle operates near a subcritical Hopf bifurcation, but never does so near a supercritical Hopf bifurcation. The similarities between simulations and observations allow us to conclude that a hair bundle possess a line of supercritical Hopf bifurcations in the high-stiffness regime

The agreement between the metrics for simulations and observations also implies that a hair bundle manifests lines of either subcritical Hopf or SNIC bifurcations at low stiffnesses. Experimentally observed bundles exhibit the graded frequency response that occurs near a SNIC but not a subcritical Hopf bifurcation. However, a model bundle near a subcritical Hopf bifurcation in the presence of noise also exhibits a graded frequency response resembling that of a SNIC bifurcation (21, 22), even though no SNIC bifurcation occurs in this region of the state diagram (16).

Graded frequency responses can arise from fluctuations inducing a system to cross a threshold. Moving a control parameter in a specific direction can increase the probability of crossing the threshold and consequently elevate the spiking frequency. Near a SNIC bifurcation, noise can cause a system to repeatedly cross a threshold and produce a spike. When operating in the low-stiffness regime, the model bundle instead possesses a quasi-threshold within the quiescent region near a subcritical Hopf bifurcation (Figs. S15-S18). In contrast with a true threshold that separates sub- and suprathreshold regimes, a quasi-threshold constitutes a region over which the model bundle can display spikes of all amplitudes (10). Noise can induce the model bundle's trajectory to cross this quasi-threshold, which can be very narrow, eliciting an excursion resembling an all-or-none spike. Increasing the stiffness diminishes both the amplitude of spikes and the range of constant forces over which spikes can occur. These noise-induced excursions are similar to the voltage spikes or action potentials produced by neurons (6, 7, 46). For example, the Hodgkin-Huxley model possesses a quasi-threshold as does the FitzHugh-Nagumo model, a two-dimensional simplification of the Hodgkin-Huxley model that is similar in form to our qualitative hair-bundle model (16, 46, 47). In the presence of noise, the FitzHugh-Nagumo model exhibits noise-induced spikes with a frequency that depends on a control parameter. Although the graded-frequency spiking present in class 1 excitable neurons has typically been described by operation near a SNIC bifurcation (6, 7, 9), similar behavior can arise instead from model-specific dynamics near a subcritical Hopf bifurcation. Some behaviors therefore do not result solely from operation near a bifurcation, but are specific to the system in question. The presence of a graded frequency response is not sufficient to conclude that a system operates near a SNIC bifurcation.

The appearance of large-amplitude oscillations as the constant force is changed is consistent with a hair bundle's crossing either a subcritical Hopf or a SNIC bifurcation. The emergence of large oscillations could alternatively arise from a third mechanism. Here a small-amplitude limit cycle is created at a supercritical Hopf bifurcation, but the amplitude of the cycle grows rapidly within an exponentially small range of control parameter values, resulting in a large-amplitude limit cycle. This phenomenon, termed a canard explosion, has been observed in a model of hair-bundle motility over a limited range of operating points (48, 49). Although canard explosions can in principle emerge in our model, it is unlikely that we observed this phenomenon in experiments. A canard explosion emerges as a sharp rise in the amplitude of oscillation with a corresponding decrease in frequency (50). In contrast, we found that large bundle oscillations appeared with a corresponding increase in frequency as the constant force declined.

Although we could not reliably distinguish a noisy system operating near a SNIC bifurcation from one poised close to a subcritical Hopf bifurcation, one could in principle discriminate between these bifurcations by assessing their phase portraits. The phase portrait of a system operating in the region of coexistence between a limit cycle and a fixed point near a subcritical Hopf bifurcation evidences a stable fixed point within a stable limit cycle. The phase portrait of a system near a SNIC bifurcation does not illustrate such coexistence. Using the analytic distributions as a proxy for the bundle's phase portraits, we at times found a region of increased probability within a loop as would be expected for a stable fixed point within a limit cycle (Supporting Material Section F.2). Although we require additional data to confirm that the bundle operates in a coexistence region, these results indicate that a bundle subjected to a small load stiffness can operate close to a subcritical Hopf bifurcation rather than a SNIC bifurcation.

Because a single model explains the behavior of bundles for both large and small values of load stiffness as arising in part from operation near supercritical and subcritical Hopf bifurcations, respectively, it is more likely that a bundle experiences subcritical Hopf rather than SNIC bifurcations. Taken together, these data indicate a clear distinction between a bundle's operation near a supercritical Hopf bifurcation at high stiffness values and its operation close to a subcritical Hopf bifurcation at low stiffnesses.

**Hair-bundle function**

Systems operating near different bifurcations exhibit distinct behaviors. Both the proximity to and type of bifurcation can determine how a system responds to different classes of stimuli. For example, a system poised near a supercritical Hopf bifurcation responds well to periodic stimuli, whereas one operating near a SNIC bifurcation can exhibit a graded frequency response in response to changes in its control parameter. Dynamical-systems analysis therefore reveals how a system might possess different functions within different regions of its state diagram. By noting where bifurcations lie in the state diagram, one can predict how that system might function in various contexts.

The stiffness of a hair bundle's load *in vivo* depends on the sensory organ in which it resides. For example, a mammalian auditory hair bundle tuned to 14 kHz might experience a stiffness load by the tectorial membrane of over 200 mN·m$^{-1}$ (43), whereas many vestibular hair bundles are coupled to otolithic membranes with a much smaller stiffness of about 1 mN·m$^{-1}$ (51). A bundle's load stiffness might therefore determine its sensory role.

When an auditory receptor organ imposes a high stiffness on a hair bundle, the bundle operates in the vicinity of a supercritical Hopf bifurcation. Under these conditions the bundle responds to periodic stimuli with robust amplification, sharp frequency selectivity, and a broad dynamic range (52, 53). If a bundle is instead coupled to a load of low stiffness, as might be the case in a vestibular organ, it operates in the vicinity of a subcritical Hopf bifurcation. Thermal fluctuations can induce spikes that permit the bundle to represent changes in constant force as changes in spike frequency. This graded frequency response could be useful for the detection of accelerations and gravistatic deflections. For operating points in the same region of a bundle's state diagram, a bundle can also spike in response to the beginning or end of a force step and can thus serve as a step detector owing to the quasi-threshold behavior that allows it to detect abrupt changes in force (16, 20).

The dual sensory roles of individual hair bundles might be mirrored by different subpopulations of the afferent neurons that innervate them. Within vestibular organs such as the sacculus and utriculus, hair cells are innervated by afferents that discharge regularly or irregularly in the absence of stimulation (54). These neurons are classified according to their distribution of interpeak time intervals, in which regular and irregular afferents possess distributions with respectively small and large coefficients of variation. Regular afferents generate action potentials with a frequency that depends on the magnitude of a constant injected current (55), a behavior resembling that of a noisy hair bundle subjected to a low stiffness. Irregular afferents, on the other hand, respond better to periodic stimuli and display spike rates that change little with the injected current (56), similar to the behavior of an oscillating hair bundle operating in the high-stiffness regime in response to changes in constant force. Whether these neuronal subpopulations operate in the vicinity of respectively subcritical or supercritical

Hopf bifurcations remains to be seen. For example, as the firing rate of a regular afferent neuron decreases, its coefficient of variation correspondingly rises (55, 57). We observe the same negative correlation between the frequency and coefficient of variation as a system crosses a bifurcation indicating that the afferent neurons might also cross a bifurcation as their control parameter is changed. The algorithms presented here may permit identification of the bifurcations near which the neurons operate. Furthermore, it remains uncertain whether these two neuronal subpopulations selectively innervate hair cells with bundles operating within different functional regimes and whether afferent neurons in the auditory system possess traits similar to those of bundles operating in the high-stiffness regime.

In summary, one sensory function—the detection of periodic stimuli—arises from a hair bundle's operation near a supercritical Hopf bifurcation. Two other sensory functions—the measurement of constant forces and the detection of force steps—result from a bundle's operation within the quiescent region near a subcritical Hopf bifurcation. These results highlight the remarkable flexibility of the hair bundle as a signal detector and suggest how the bundle might have evolved through changes in its operating point to serve disparate functions in various auditory, vestibular, and lateral-line organs.

## Acknowledgments

We thank the members of our research group for comments on the manuscript. JDS is supported by grants F30DC013468 and T32GM07739 from the National Institutes of Health. AJH is an Investigator of Howard Hughes Medical Institute.

## Authors' Contributions

JDS, DÓM, and AJH designed the experiments. JDS performed the experiments and the simulations. JDS, DÓM, and AJH analyzed the data and wrote the manuscript. The authors declare no competing financial interests.

**Supporting References**

3434

3467. Kralemann, B., L. Cimponeriu, M. Rosenblum, A. Pikovsky, and R. Mrowka. Phase dynamics of coupled oscillators reconstructed from data. Physical Review E, 77(066205):1-16, 2008.

68. Zhu, Y., Y.-H. Hsieh, R. R. Dhingra, T. E. Dick, F. J. Jacono, and R. F. Galán. Quantifying interactions between real oscillators with information theory and phase models: Application to cardiorespiratory coupling. Physical Review E, 87(2):022709, 2013.

67. Kralemann, B., L. Cimponeriu, M. Rosenblum, A. Pikovsky, and R. Mrowka. Phase dynamics of coupled oscillators reconstructed from data. Physical Review E, 77(066205):1-16, 2008.

68. Zhu, Y., Y.-H. Hsieh, R. R. Dhingra, T. E. Dick, F. J. Jacono, and R. F. Galán. Quantifying interactions between real oscillators with information theory and phase models: Application to cardiorespiratory coupling. Physical Review E, 87(2):022709, 2013.

# Figure Legends

**Figure 1. The hair bundle's state diagram.** (A) A theoretical state diagram specifies the behaviors of a hair bundle for various combinations of constant force ($F_C$) and load stiffness ($K_L$). A line of Hopf bifurcations (red) separates a region within which a hair bundle oscillates spontaneously (orange) from one in which the bundle remains quiescent (white). Bautin points (black) lie at the border between the supercritical and subcritical segments of this line. Within the spontaneously oscillatory region, a bundle's oscillations fall in amplitude and rise in frequency with an increase in load stiffness (red and blue arrows, right panels). A line of fold bifurcations (green, dashed) confines a region within which a bundle exhibits bistability (light green). The gray arrows correspond to the regions of the bundle's state diagram explored in Figures 2-5. (B) Oscillations of an experimentally observed hair bundle changed in appearance as the load stiffness increased from 300 µN·m$^{-1}$ to 1000 µN·m$^{-1}$ (red to purple). The position histograms and associated dip statistics shown to the right of each trace signal the presence or absence of spontaneous oscillations. The dip statistic quantifies the modality of the system's position histogram; a dip statistic of at least 0.01 with $p < 10^{-3}$ implies spontaneous oscillation. (C) An experimental state diagram depicts the behavior of a hair bundle for different values of the load stiffness and constant force. White boxes correspond to operating points at which the bundle's position histogram possessed a dip statistic less than 0.01, indicating quiescent behavior. Within the territory of spontaneous oscillation, the intensities of red and blue correspond to respectively the amplitude and frequency of the bundle's oscillations calculated with a peak-detection algorithm with a threshold of 25 nm. Colored circles correspond to the traces in (B). (D) The dip-statistic values for the experimental state diagram in (C) are displayed in shades of purple. (E) Spearman's rank correlation (ρ) quantifies the relationships between amplitude (red arrow) and frequency (blue arrow) with the load stiffness (ρ($K_L$)) and constant force (ρ($F_C$)) (Table S1). For all experimental data, the stiffness and drag coefficient of the stimulus fiber were respectively $k_{SF}$ = 109 µN·m$^{-1}$ and $\xi_{SF}$ = 142 nN·s·m$^{-1}$. The proportional gain of the load clamp was 0.01. For a complete description of the bundle's experimental state diagram, see Supporting Material Section B.

**Figure 2. Crossing a supercritical Hopf bifurcation by adjusting the constant force.** Over a range of parameter values, we compared behavior near a supercritical Hopf bifurcation with additive noise (A-G) to that of an experimentally observed hair bundle (H-N). (A) A noisy system operating near a supercritical Hopf bifurcation displayed noise-induced ringing when the control parameter was negative but oscillated spontaneously for positive values (left). When the amplitude of spontaneous oscillation surpassed that of the noise, the position histogram (middle) disclosed two peaks. When the control parameter exceeded zero, the joint probability distribution (right) for each real-valued position ($X$) and its Hilbert transform ($X_H$)—the analytic distribution—was circular, with a diameter that grew with the control parameter. In all analytic distributions, red corresponds to high and blue to low probability values. Probabilities at or near zero are displayed in white. (B) The dip statistic for the position distribution is shown as function of the control parameter. A low value for the dip statistic corresponded to a unimodal distribution (green). When the control parameter exceeded zero and the amplitude of oscillation exceeded the level of the noise, the position histogram displayed two peaks (purple) corresponding to a large value of the dip statistic. The control parameter value at which the dip statistic achieved significance (shaded; $p < 10^{-3}$) served as an estimate of the bifurcation's location (dashed line). (C) The RMS magnitude of the system's motion rose gradually as the control parameter increased. (D) The amplitude rose with the control parameter for peak-detection thresholds of 0.68 (red), 1.06 (yellow), and 1.45 (cyan). (E) The frequency of oscillation, with the same thresholds as in (D), grew with an increase in the control parameter until it achieved a constant value. The value of the control parameter at which the rise in frequency was steepest increased with the threshold. (F) The coefficient of variation describes the regularity of oscillation, with a high value corresponding to increased variability in the interpeak time interval. The value of the control parameter at which the coefficient of variation crossed 0.5 increased with the peak-detection threshold. (G) The analytic information rose gradually with the control parameter. (H) The behavior of an experimentally observed hair bundle was transformed by changes in the constant force: as the force decreased, the amplitude of spontaneous oscillation rose (left) and the position histogram became bimodal (middle). The analytic distribution formed a loop whose diameter increased with a decrease in the constant force (right). (I) Reflecting a unimodal position histogram, the dip statistic remained small for large values of the constant force (green). When the constant force fell below 20 pN, the position histogram displayed two peaks (purple) and the dip statistic increased, defining the boundary of an oscillatory regime (shaded; $p < 10^{-3}$). (J) The RMS magnitude of the bundle's motion rose gradually as the constant force decreased. (K) The amplitude increased as the constant force declined for peak-detection thresholds of 6 nm (red), 8 nm (yellow), and 10 nm (cyan). (L) The frequency of oscillation rose gradually with a decrease in constant force until it achieved a constant value near 5 pN. The force at which this rise in frequency occurred depended on the threshold value. (M) The coefficient of variation exceeded 0.5 at a value of constant force that decreased as the peak-detection threshold increased. (N) The analytic information rose gradually as the constant force fell. For all experimental data, the load stiffness was 400 $\mu$N·m$^{-1}$ with a gain of 0.1. The stiffness and drag coefficient of the stimulus fiber were respectively 109 $\mu$N·m$^{-1}$ and 142 nN·s·m$^{-1}$. All simulations possessed noise with standard deviations of $\sigma_R = \sigma_I = 0.1$. The error bars represent the standard errors of the means.

**Figure 3. Traversing a supercritical Hopf bifurcation by altering the load stiffness.** Over a range of load stiffnesses, we compared the behavior of a model hair bundle crossing a supercritical Hopf bifurcation (A-G) to that of an experimentally observed hair bundle (H-N). All of the panels depicted here display the same metrics as shown in Figure 2. (A,H) As the stiffness

declined, the amplitude of spontaneous oscillation rose (left), the position histogram became bimodal (middle), and the diameter of the limit cycle increased (right). (B,I) The dip statistic rose as the stiffness decreased, signalling the emergence of bimodal position histograms and the onset of spontaneous oscillations (shaded; $p < 10^{-3}$). The estimated boundary of the oscillatory regime occurred on the oscillatory side of the deterministic bifurcation (dashed line). (C,J) The RMS magnitude rose gradually as the stiffness decreased. (D,K) The amplitude increased with a fall in stiffness. For these and subsequent panels, we used peak-detection thresholds of (D-F) 1 (red) and 1.5 (cyan) or of (K-M) 26 nm (red) and 34 nm (cyan). (E,L) The stiffness at which the frequency rose depended on the value of the peak-detection threshold. Spontaneous oscillations emerged with a frequency that first rose and subsequently fell with decreases in stiffness. (F,M) The stiffness at which the coefficient of variation exceeded 0.5 depended on the threshold value. (G,N) The analytic information rose gradually as the stiffness decreased. For all panels, the constant force was zero with a gain of 0.1. The stiffness and drag coefficient of the stimulus fiber were respectively 109 $\mu N \cdot m^{-1}$ and 142 $nN \cdot s \cdot m^{-1}$. Dashed lines correspond to the location of the supercritical Hopf bifurcation in the absence of noise. Stochastic simulations employed noise with standard deviations of $\sigma_x = \sigma_f = 0.1$. The error bars represent the standard errors of the means.

**Figure 4. A subcritical Hopf or SNIC bifurcation at low stiffness.** We compared the behavior of noisy systems near either a subcritical Hopf bifurcation (A-G) or a SNIC bifurcation (H-N) with that of an experimentally observed hair bundle subjected to changes in constant force (O-U). Each row of panels describes the same metrics as the corresponding rows in Figures 2 and 3. (A) A noisy system poised near a subcritical Hopf bifurcation displayed noise-induced bursts of high-amplitude oscillations when the control parameter fell between -0.25 and 0, a regime in which a limit cycle coexisted with a stable fixed point (left). When the control parameter exceeded zero the system exhibited high-amplitude sinusoidal oscillations. The position histogram featured multiple peaks when the control parameter exceeded -0.25, a regime bounded by a saddle-node of limit cycles and a subcritical Hopf bifurcation (middle). The analytic distribution formed a cycle whose diameter changed little with the control parameter and in some cases included a stable region at the center (right). (B) The dip statistic evidenced the oscillatory boundary within the coexistence region of a subcritical Hopf bifurcation (shaded; $p < 10^{-3}$). (C) The RMS magnitude grew gradually with the control parameter. (D) The amplitude rose abruptly with the control parameter for peak-detection thresholds of 1.2 (red), 1.4 (yellow), and 1.6 (cyan). (E) The frequency of oscillation rose briskly between the saddle-node of limit cycles and the subcritical Hopf bifurcation and then gradually on the exclusively oscillatory side of the Hopf bifurcation. (F) The coefficient of variation was insensitive to changes in the peak-detection threshold. (G) The analytic information rose sharply near the subcritical Hopf bifurcation. (H) A system poised near a SNIC bifurcation displayed large-amplitude spikes that increased in frequency with a rising control parameter (left). In agreement with this behavior, the position histograms became increasingly bimodal (middle). The analytic distribution disclosed a cycle whose diameter remained invariant to changes in the control parameter (right). Consistent with its spiking behavior, the system dwelled more often along one part of the cycle for low values of the control parameter. (I) The dip statistic determined the location of the deterministic SNIC bifurcation (shaded; $p < 10^{-3}$). (J) The RMS magnitude grew rapidly to a constant value near the SNIC bifurcation. (K) The amplitude remained constant on the oscillatory side of the SNIC bifurcation for thresholds of 0.7 (red), 0.9 (yellow), and 1.1 (cyan). (L) The frequency of oscillation rose gradually from zero near the SNIC bifurcation for all peak-detection thresholds. (M) The coefficient of variation was insensitive to the peak-detection threshold near the SNIC bifurcation. (N) The analytic information reached a plateau for positive values of the control parameter. (O) As the constant force decreased, the spontaneous oscillations of an experimentally observed hair bundle resembled spikes of rising frequency (left), the position histograms became increasingly bimodal (middle), and the analytic distributions formed loops

that changed little in diameter with a locus of greater probability along one section of each cycle (right). (P) The dip statistic located the boundary of the spontaneously oscillatory region (shaded; $p < 10^{-3}$). (Q) The bundle's RMS magnitude rose abruptly to a nearly constant value near the oscillatory boundary. (R) The amplitude remained nearly constant for constant forces below 25 pN for peak-detection thresholds of 20 nm (red), 27.5 nm (yellow), and 35 nm (cyan). (S) The bundle's frequency of spontaneous oscillation rose gradually from zero for constant forces below 30-35 pN. (T) The coefficient of variation remained insensitive to changes in threshold. (U) The analytic information rose sharply and then gradually as the constant force decreased. For all experimental data, the load stiffness was 50 $\mu$N·m$^{-1}$ with a gain of 0.1. The stiffness and drag coefficient of the stimulus fiber were respectively 139 $\mu$N·m$^{-1}$ and 239 nN·s·m$^{-1}$. Black dashed lines in (B-G) and (I-N) correspond respectively to the locations of a subcritical Hopf bifurcation and a SNIC bifurcation. Pink dashed lines depict the location of a saddle-node of limit cycles bifurcation. Stochastic simulations for a subcritical Hopf bifurcation and for a SNIC bifurcation possessed noise levels of $\sigma_R = \sigma_I = 0.2$. The error bars represent the standard errors of the means.

**Figure 5. A subcritical Hopf bifurcation can be crossed by controlling the constant force.** When the load stiffness remains low, a decrease in constant force advances a bundle's operating point across a subcritical Hopf bifurcation. Stochastic simulations of a model of hair-bundle motility (A-G) were compared with an experimentally observed hair bundle (H-N). The results depicted here correspond to the same metrics displayed in Figures 2-4. (A) As the constant force decreased, a model bundle exhibited noise-induced spikes of rising frequency (left), an increasingly bimodal position histogram (middle), and an analytic distribution with a cycle whose diameter changed little over a range of forces and upon which rested a locus of higher probability (right). (B) The dip statistic defined an oscillatory boundary at a control parameter smaller than the control parameters corresponding to the deterministic bifurcations (shaded; $p < 10^{-3}$). (C) The model bundle's RMS magnitude rose to a nearly constant value as the constant

force decreased. (D) Calculated with thresholds of 1 (red) and 2 (cyan), the amplitude of the bundle's motion remained constant for forces below 1. (E) The frequency of oscillation for a model bundle rose smoothly from zero as the constant force decreased for both peak-detection thresholds. (F) The coefficient of variation remained insensitive to changes in the peak-detection threshold. (G) The analytic information rose with a decrease in constant force. (H) An experimentally observed hair bundle exhibited behaviors that accorded with those of a model bundle crossing a subcritical Hopf bifurcation in the presence of noise. As the constant force declined, the bundle displayed spikes of increasing frequency (left), a more clearly bimodal position histogram (middle), and an analytic distribution with a cycle that changed little in diameter over a range of forces and upon which rested a locus of enhanced probability (right). (I) The dip statistic defined the boundary of the oscillatory region (shaded; $p < 10^{-3}$). (J) The RMS magnitude of the bundle's motion rose sharply as the constant force fell below 40-45 pN and achieved a constant value for forces below 35 pN. (K) The amplitude of the bundle's oscillations achieved a nearly constant value for forces below 35-48 pN. We employed peak-detection thresholds of 50 nm (red) and 60 nm (cyan). (L) The bundle's frequency of oscillation rose gradually from zero as the constant force decreased. (M) The constant force at which the coefficient of variation exceeded 0.5 remained relatively insensitive to changes in the peak-detection threshold. (N) The analytic information rose as the constant force decreased. For experimental data, the load stiffness was 100 $\mu N \cdot m^{-1}$ with a gain of 0.1. The stiffness and drag coefficient of the stimulus fiber were respectively 139 $\mu N \cdot m^{-1}$ and 239 $nN \cdot s \cdot m^{-1}$. Black dashed lines correspond to the location of the subcritical Hopf bifurcation at $F_C = 0.66$ and pink dashed lines to the location of the saddle-node of limit cycles bifurcation at $F_C = 0.664$ in the absence of noise. Stochastic simulations of the model of hair-bundle motility possessed a stiffness of 2 and noise levels of $\sigma_x = \sigma_f = 0.2$. The error bars represent standard errors of the means.

**Figure 6. Summary of metrics for the identification of bifurcations.** We present schematic diagrams for each of the metrics used to identify the type of the bifurcation near which a system operates. The three columns on the left include diagrams for noisy systems crossing supercritical Hopf, subcritical Hopf, and SNIC bifurcations. The two columns on the right refer to a model hair bundle crossing either a supercritical Hopf (red) or a subcritical Hopf (blue) bifurcation. Red and blue arrows highlight the range of parameter values explored in each instance. Green arrows indicate trends in each of the statistics as a function of the control parameter. Purple arrows and purple arrowheads illustrate respectively the dependence and independence of the frequency of oscillation and coefficient of variation with changes in peak-detection threshold. For each of the panels, we highlight in color the metrics that agree with observations of bundles subjected to high- (red) and low-stiffness loads (blue).

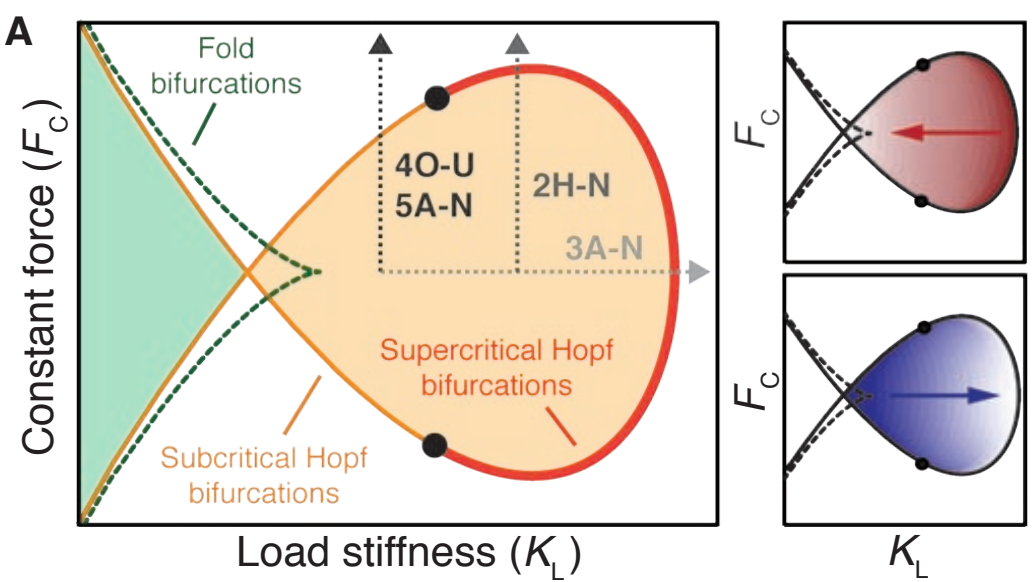
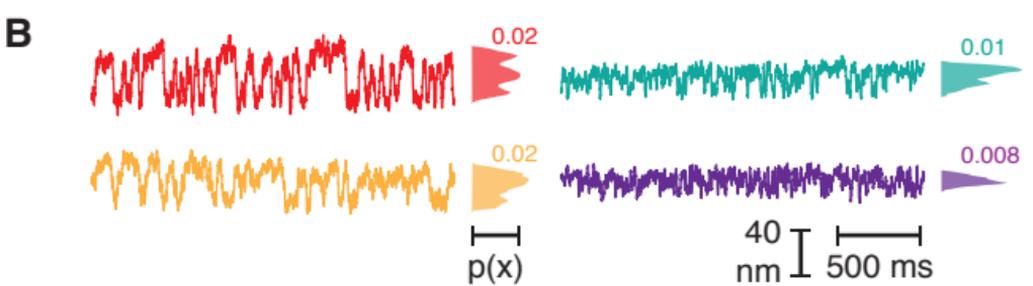
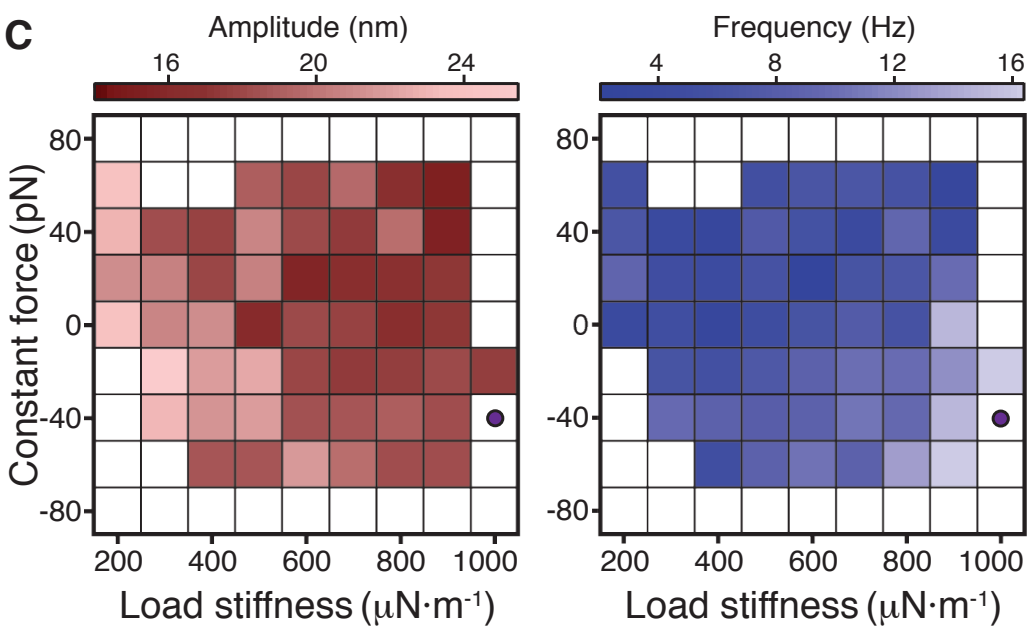
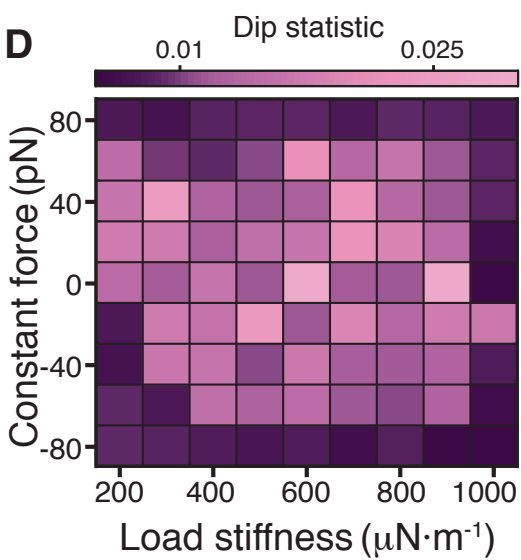
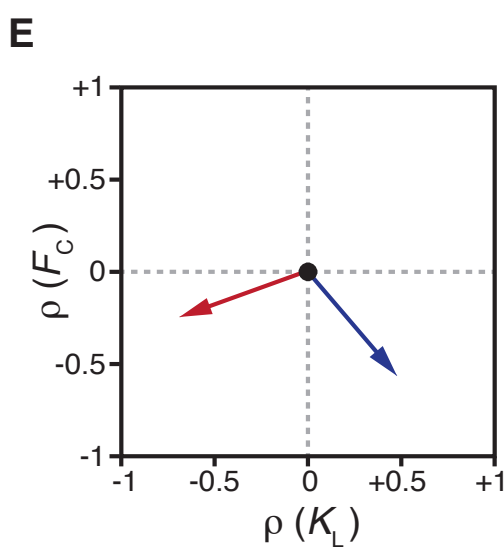

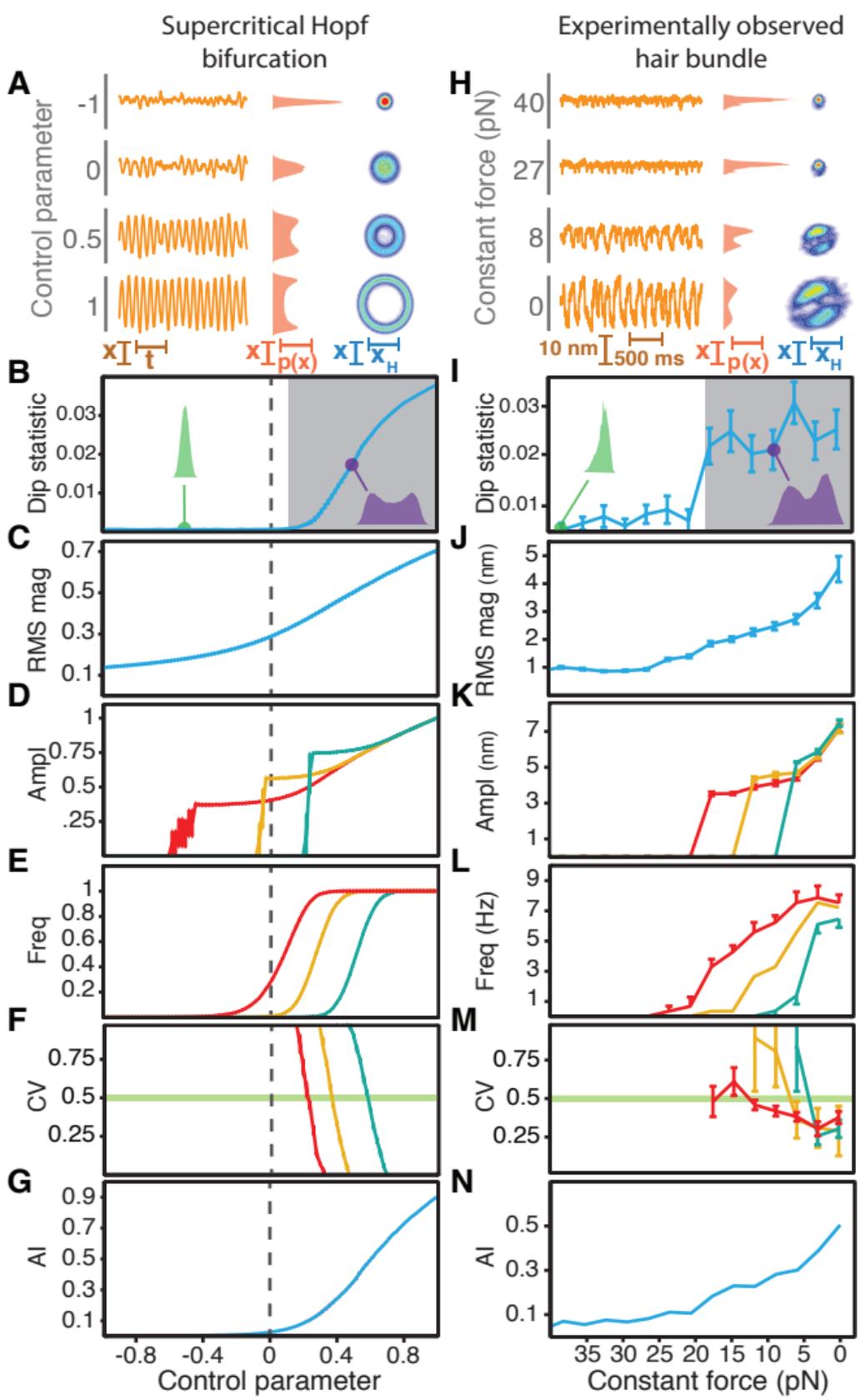

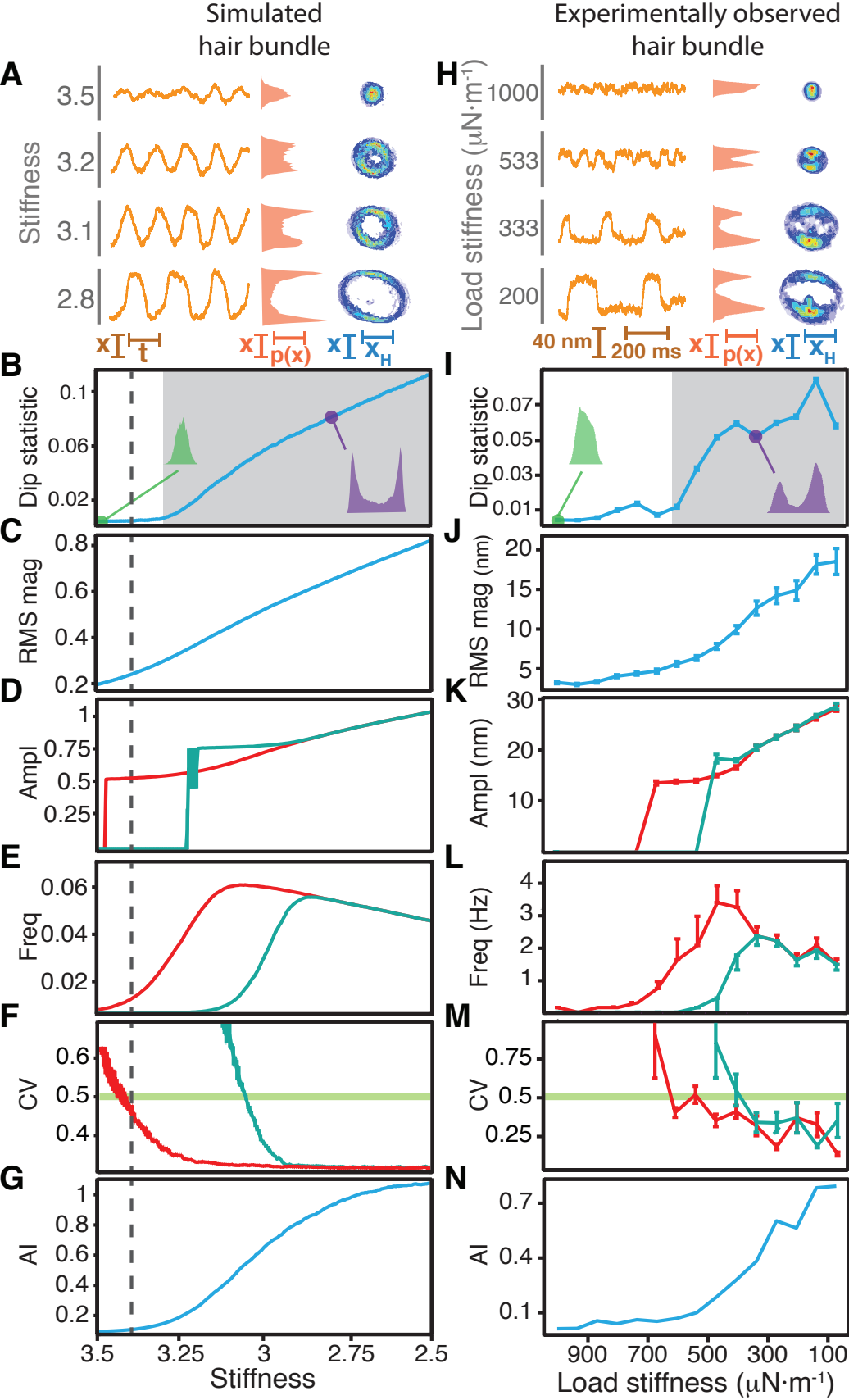

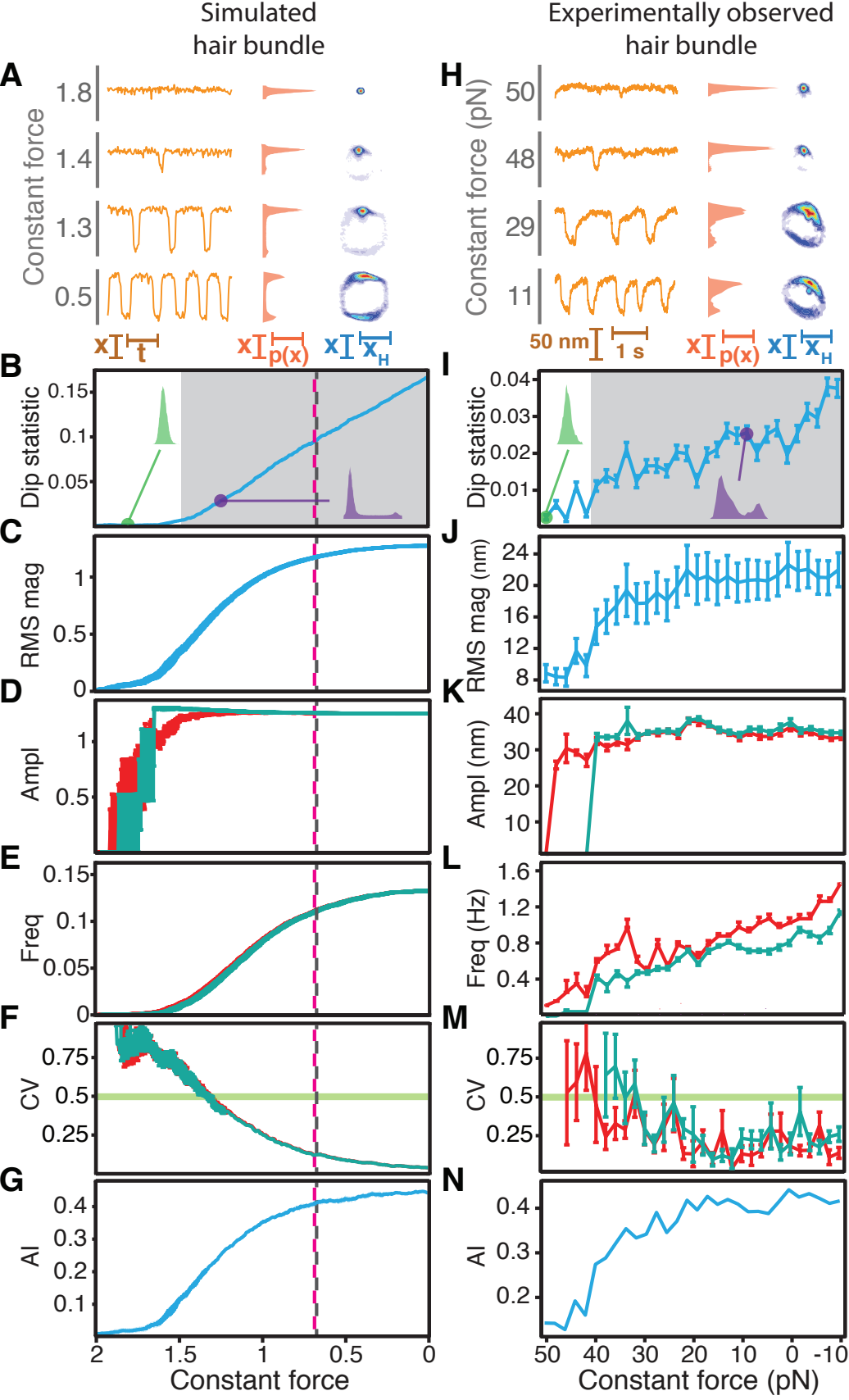

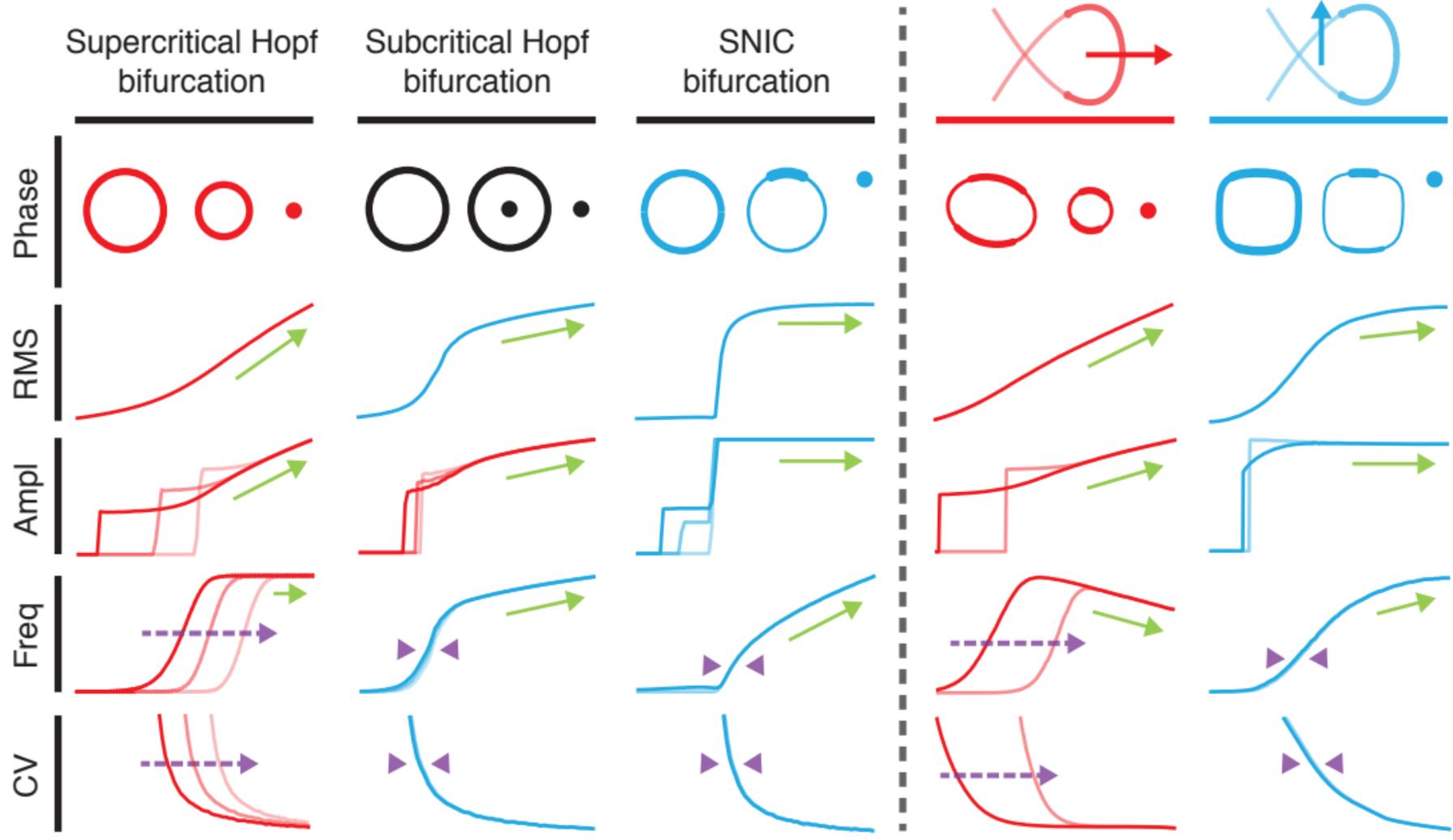